\documentclass{aastex62}

\newcommand{\beq}{\begin{equation}}
\newcommand{\eeq}{\end{equation}}
\newcommand{\inttau}{\int \tau_{\rm HI} d{\rm V}}
\newcommand{\zpk}{z_{\rm peak}}

\newcommand{\ts}{{\rm T_s}}

\newcommand{\cm}{cm$^{-2}$}

\newcommand{\kmps}{km~s$^{-1}$}

\newcommand{\nhi}{\rm{N_{H{\small \textsc i}}}}

\newcommand{\hi}{H{\sc i}}
\newcommand{\sigc}{\sigma_{\rm cont}}
\newcommand{\sigl}{\sigma_{\rm HI}}
\newcommand{\hii}{H{\sc i} 21\,cm}
\defcitealias{NatPaper}{vD18}
\graphicspath{{./}{figures/}}

\shorttitle{High-opacity H{\sc i} 21\,cm absorbers at $z \approx 1.2$}
\shortauthors{Chowdhury et al.}

\begin{document}

\title{Giant Metrewave Radio Telescope detections of two high-opacity H{\sc i} 21\,cm absorbers at $z \approx 1.2$}

\correspondingauthor{Aditya Chowdhury}
\email{chowdhury@ncra.tifr.res.in}

\author{Aditya Chowdhury}
\affil{National Centre for Radio Astrophysics,\\ Tata Institute of Fundamental Research, \\ Pune 411007, India.}

\author{Nissim Kanekar}
\affil{National Centre for Radio Astrophysics,\\ Tata Institute of Fundamental Research, \\ Pune 411007, India.}

\author{Jayaram N. Chengalur}
\affil{National Centre for Radio Astrophysics,\\ Tata Institute of Fundamental Research, \\ Pune 411007, India.}



\begin{abstract}

We report the discovery of two remarkable high-opacity H{\sc i} 21\,cm absorbers against low-luminosity active galactic nuclei (AGNs), at $z = 1.2166$ towards J0229+0044 and at $z=1.1630$ towards J0229+0053. The absorbers were detected in an unbiased  Giant Metrewave Radio Telescope survey for H{\sc i} 21\,cm absorption against radio sources in the DEEP2 survey fields, covering $z \approx 0.73-1.53$, and including sources without known redshifts. The velocity-integrated H{\sc i} 21\,cm optical depths are $(74.2 \pm 7.8)$~km~s$^{-1}$ (J0229+0044) and $(78.41 \pm 0.81)$~km~s$^{-1}$ (J0229+0053), higher than that of any known redshifted H{\sc i} 21\,cm absorber at $z > 0.12$, and implying high H{\sc i} column densities, $> 10^{22}$~cm$^{-2}$. The emission redshift of J0229+0044 is consistent with the H{\sc i} 21\,cm absorption redshift, while the strength and velocity spread of the absorption against J0229+0053 suggest that it too arises from gas in the AGN environment: both absorbers are thus likely to be ``associated'' systems. The two AGNs have low rest-frame 1.4~GHz radio and 1215~\AA\ ultraviolet luminosities ($\lesssim 10^{26.1}$~W~Hz$^{-1}$ and $\lesssim 10^{21.7}$~W~Hz$^{-1}$, respectively), both significantly lower than the typical luminosities of AGNs against which H{\sc i} 21\,cm searches have hitherto been carried out at $z \gtrsim 1$. The paucity of H{\sc i} 21\,cm absorbers at $z \gtrsim 1$ may be due to a luminosity bias in high-$z$ AGN samples that have been searched for H{\sc i} 21\,cm absorption, where the high AGN ultraviolet luminosity affects physical conditions in its environment, ionizing the neutral hydrogen.

\end{abstract}


\section{Introduction}

For radio-loud active galactic nuclei (AGNs), studies of ``associated'' \hii\ absorption by neutral 
atomic hydrogen (\hi) in the AGN environment provide an interesting probe of physical conditions in the 
vicinity of AGNs  \citep[see][for a recent review]{Morganti18}. 
For example, the strength of the \hii\ absorption yields information on the \hi\ column density $\nhi$ and the gas 
spin temperature $\ts$, and how these are affected by proximity to the AGN. The \hii\ absorption kinematics, 
relative to the AGN, can be used to determine whether the \hi\ is infalling (redshifted) or outflowing 
(blueshifted), and thus, to infer the importance of AGN fuelling and feedback at different cosmological 
epochs \citep[e.g.][]{Gorkom89,Vermeulen03,Morganti03,Morganti16}. Very long baseline interferometric \hii\ absorption studies can be used to track the influence of the AGN jets on the gas, and the driving of high-velocity outflows \citep[e.g.][]{Oosterloo00,Morganti13}. The detection rates of \hii\ absorption in different AGN types can provide information on the evolutionary history of AGNs, and on AGN unification schemes \citep[e.g.][]{Vermeulen03}.
High-opacity \hii\ absorbers are also good candidates for searches for radio molecular absorption 
\citep[e.g.][]{Wiklind94,Kanekar05,Allison19}, which can be used to probe fundamental constant evolution on cosmological 
timescales \citep[e.g.][]{Kanekar11,Kanekar18}.





The detection of associated \hii\ absorption at high redshifts, $z > 1$, would allow one to extend the above studies to the early Universe and track the cosmological evolution of \hi\ in AGN environments. Unfortunately, despite a number of searches \citep[e.g.][]{Curran08,Aditya18gps,Aditya18cjf,Grasha19}, this endeavour remains limited by the paucity of known high-$z$ associated \hii\ absorbers. While more than a hundred such absorbers have been found at $z < 1$ \citep[e.g.][]{Vermeulen03,Gereb15,Maccagni17,Aditya19}, there are only 8 confirmed detections at $z>1$ \citep{Uson91,Moore99,Ishwara-Chandra03,Curran13,Aditya17,Aditya18gps,Dutta20}. {\citet{Curran08} argued that this apparent redshift evolution in the strength of the associated \hii\ absorption might arise because higher-redshift AGNs searched for \hii\ absorption typically have high ultraviolet (UV) and radio luminosities; however, their results were based on a heterogenous AGN sample.} This issue was further examined in an \hii\ absorption survey of a homogenous sample of flat-spectrum AGNs by \citet{Aditya18cjf}, who found that associated \hii\ absorption is significantly weaker in AGNs at high redshifts, but also in AGNs with high rest-frame radio or ultraviolet (UV) luminosities. Unfortunately, the luminosity bias in their target sample, with high-luminosity AGNs at high redshifts, meant that it was not possible to break the above degeneracy between redshift evolution and AGN luminosity.  Searches for associated \hii\ absorption in low-luminosity AGNs at high redshifts offer the best route to break this degeneracy. 


Next, searches for associated \hii\ absorption have so far mostly been carried out in AGNs with known redshifts. This introduces a bias against AGN environments with high \hi\ column densities, as such systems are likely to have high dust columns, which would obscure the AGN in the optical and UV wavebands, making it difficult to measure its redshift. \hii\ or mm-wave absorption surveys that target {\it all} radio AGNs independent of redshift information ensure no bias against dusty sightlines \citep[e.g.][]{Wiklind96,Kanekar14gbt,Allison15}.

We have used the upgraded Giant Metrewave Radio Telescope \citep[GMRT;][]{Gupta17} to carry out such an unbiased search for \hii\ absorption, covering the redshift range $z \approx 0.73-1.53$ against a large sample of radio sources. In this {\it Letter}, we report the discovery of two remarkable high-opacity \hii\ absorbers at $z \approx 1.2$, both identified against reddened AGNs, with low rest-frame radio and UV luminosities.

\section{Observation, Data Analysis, and Results}
\label{sec:obs}

\begin{figure*}
    \centering
    \includegraphics[width=0.489\linewidth]{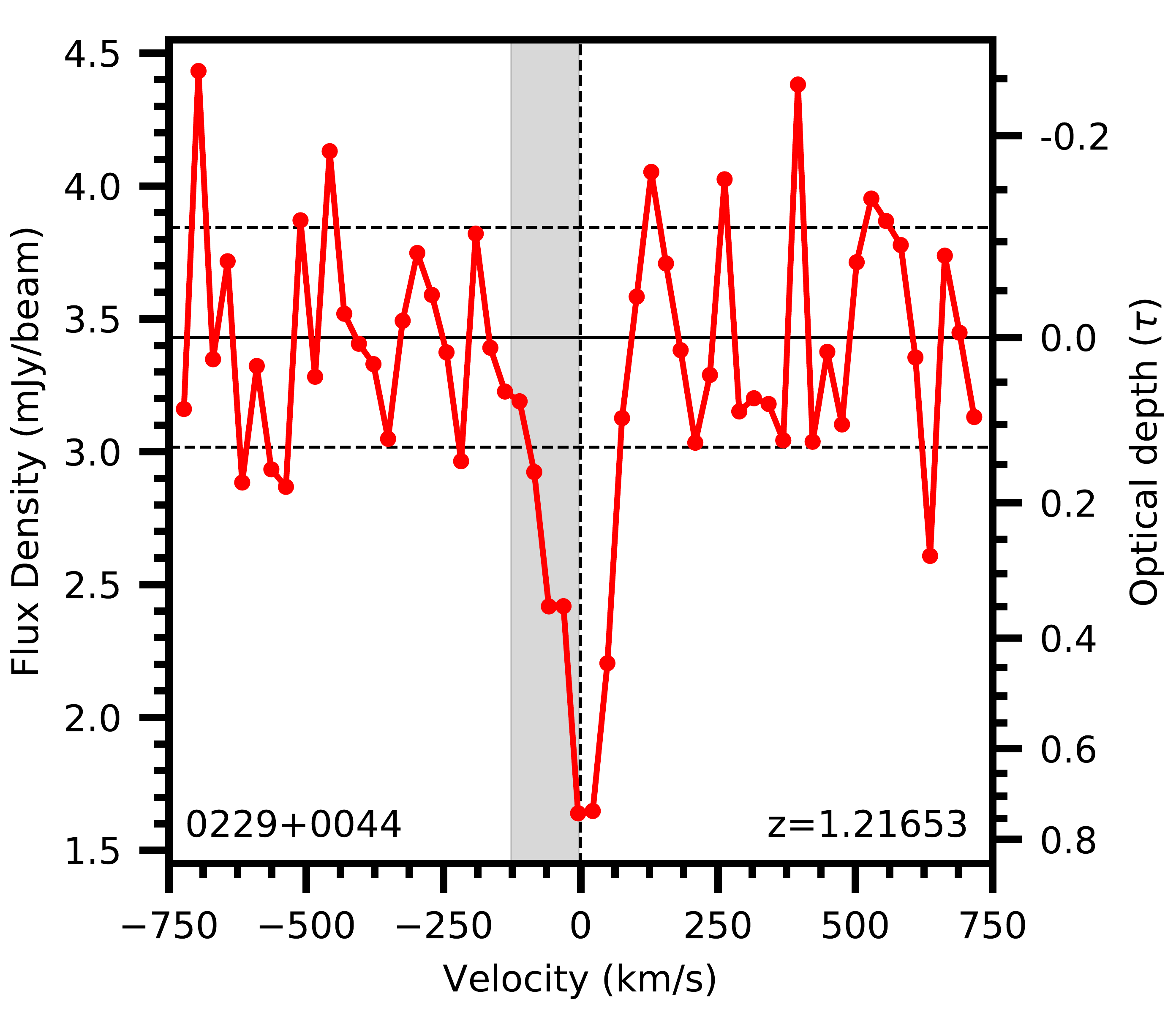}
    \hspace{0.15in}
    \includegraphics[width=0.475\linewidth]{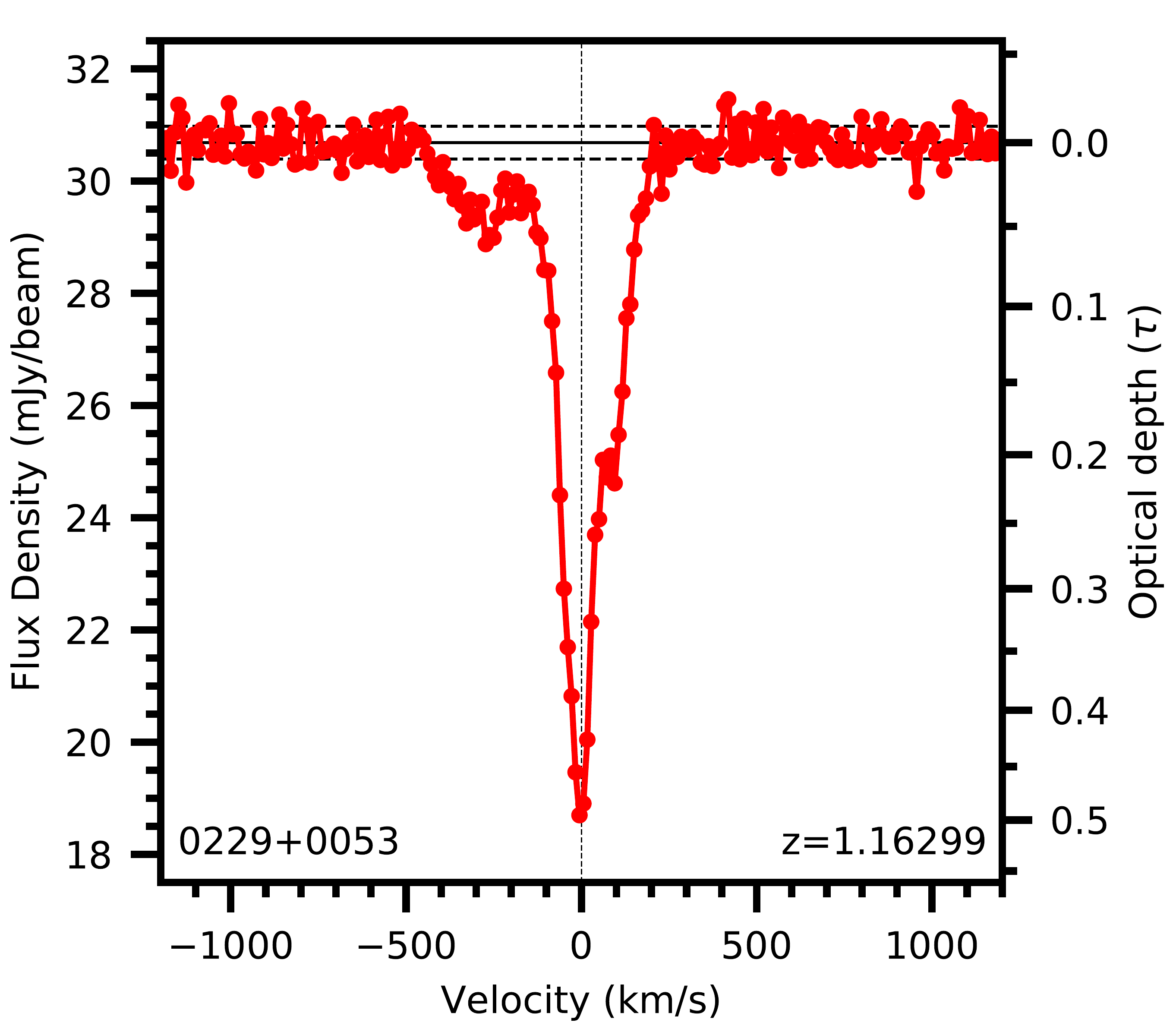}
    \caption{The \hii\ spectra towards [A]~J0229+0044 and [B]~J0229+0053, at velocity resolutions of 
$\approx 26.6$~\kmps\ and $\approx 11.2$~\kmps, respectively. The solid horizontal line in each panel
indicates the AGN continuum flux density at the redshifted \hii\ line frequency. The $1\sigma$ RMS 
noise values on the spectra (the horizontal dashed lines) are 0.41~mJy per $\approx 26.6$~\kmps\ 
(J0229+0044) and 0.29~mJy per $\approx 11.2$~\kmps\ (J0229+0053). {The velocity axes in both panels are with respect to the redshift of the \hii\ absorption.} The shaded region in the left panel 
indicates the optical spectroscopic redshift (and error) of J0229+0044, which is consistent with the 
\hii\ absorption redshift.}
    \label{fig:spectra}
\end{figure*}

{The two new \hii\ absorbers were discovered in our $550-850$~MHz survey of the DEEP2 Survey fields, using the Band-4 receivers of the upgraded GMRT. The DEEP2 Survey used the DEIMOS spectrograph on the Keck~II Telescope to accurately measure the spectroscopic redshifts of 38,000 galaxies at $z \approx 0.70-1.45$, in four regions of the sky \citep{Newman13}. Our GMRT survey used five pointings to cover two of the four DEEP2 field, with the main aim of measuring the average \hi\ mass of star-forming galaxies at $0.74 \leq z \leq 1.45$, by stacking the \hii\ emission spectra of galaxies with DEEP2 spectroscopic redshifts \citep{Chowdhury20}. However, in addition to the primary science goal,} the high spectral resolution of our survey ($\approx 18-25$~\kmps) allowed us to carry out an unbiased search for \hii\ absorption against all the ($\approx 250$) radio-continuum sources with flux densities $\geq 1$~mJy (without correcting for the shape of the primary beam) in the five target fields, including sources without a known redshift. Our search was sensitive to both associated and intervening \hii\ absorption at $ z \approx 0.73-1.53$. This search yielded the initial discovery of two high-opacity absorbers, at $z=1.2166$ towards J0229+0044, and at $z=1.1630$ towards J0229+0053, the focus of this {\it Letter}.

We followed up the two absorbers with GMRT Director's Discretionary Time (DDT) in December~2019 (J0229+0044; proposal DDTC101) and February~2020 (J0229+0053; proposal DDTC121) to confirm the \hii\ absorption, and accurately measure the \hii\ optical depth. {The total observing times were 10~hours (J0229+0044) and 12~hours (J0229+0053), using the GMRT Software Backend (GSB) and the GMRT Wideband Backend (GWB), respectively, as the primary correlator to cover the redshifted \hii\ line. The choice of  correlator and the correlator settings were based on the aim of simultaneously covering the redshifted Lambda-doubled OH~18\,cm lines; however, local oscillator tuning issues in the observations of J0229+0044 led us to a different correlator setup for the observations of J0229+0053. For J0229+0044, the OH~18\,cm line frequencies were covered using the GWB as the secondary correlator, with a bandwidth of 200~MHz, centered at 700~MHz and sub-divided into 16,384 channels. Conversely, for J0229+0053, a single GWB 200~MHz band with 16,384 channels was used to cover both the \hii\ line and the OH~18\,cm lines.} The observational details (related to the \hii\ line) for the two sources are provided in Table~\ref{table:results}.

The data were analyzed in the Common Astronomy Software Package \citep[{\sc casa} version~5; ][]{McMullin07}, following standard procedures. The {\sc aoflagger} package \citep{Offringa12} was additionally used for excision of data affected by radio frequency interference (RFI). After the initial data editing, and gain and bandpass calibration {using observations of the compact source 0204+152}, a standard self-calibration procedure was carried out for each target source, using line-free channels. 
The imaging was done with the task {\sc tclean}, with the w-projection 
algorithm \citep{Cornwell08}. In the case of J0229+0053, where the observing bandwidth was large, we additionally 
used the MT-MFS algorithm \citep[second order expansion; ][]{Rau11} to image the target field. 
The synthesized beams of the final continuum images have full widths at half maximum (FWHMs) of $6.7'' \times 5.1''$ (J0229+0044) and $5.6'' \times 4.4''$ 
(J0229+0053). Both sources were found to be unresolved in our continuum images, with flux densities of 
$(3.43 \pm 0.38)$~mJy (J0229+0044, at $\approx 640$~MHz) and $(31.08\pm 0.41)$~mJy (J0229+0053, at $\approx 657$~MHz).
The quoted errors include the error measured from a 2-dimensional Gaussian fit to a small region around each source and the typical error in the GMRT flux density scale, $\approx 10$\% at these frequencies.

The task {\sc uvsub} was then used to subtract out all detected continuum emission from the self-calibrated 
visibilities of each field. For each source, a spectral cube was made from these continuum-subtracted 
visibilities in the barycentric frame, using natural weighting. The \hii\ spectrum was extracted by taking a 
cut through the cube at the AGN location. The final spectrum for each source was then obtained by fitting a 
second-order polynomial to the line-free regions of each spectrum, and subtracting this out. 
 We further smoothed the spectra by 7~channels (J0229+0044) and 2~channels (J0229+0053), to velocity resolutions of $\approx 26.6$~\kmps\ and $\approx 11.2$~\kmps, respectively. The final optical depth RMS noise values are $\approx 0.12$ per 26.6~\kmps\ (J0229+0044) and $0.0068$ per 11.2~\kmps\ (J0229+0053).

\begin{table*}[t!]
\centering
\caption{\hii\ observational details and results. The columns are (1)~the AGN name, (2)~the J2000 co-ordinates, (3)~the 
bandwidth, BW, in MHz, covering the \hii\ line, (4)~the redshifted \hii\ line frequency, $\nu_{\rm HI}$, in MHz, (5), the RMS noise on the 
continuum image, in $\mu$Jy/Bm, (6)~the AGN flux density at $\nu_{\rm HI}$, in mJy, (7)~the original spectral 
resolution, $\Delta V$, in \kmps, (8)~the RMS noise at the resolution $\Delta V$, in mJy, (9)~the velocity-integrated 
\hii\ optical depth, $\inttau$, in \kmps, (10)~the \hi\ column density, $\nhi$, assuming $\ts = 100$~K and $f = 1$, and 
(11)~the redshift of the peak of the \hii\ optical depth, $\zpk$.
\label{table:results}}
\begin{tabular}{|c|c|c|c|c|c|c|c|c|c|c|}
\hline
AGN        & Coordinates  & BW   & $\nu_{\rm HI}$ & $\sigc$    &  S               & $\Delta V$ & $\sigl$ & $\inttau$        & $\nhi$               & $\zpk$ \\
	   &   (J2000)    & MHz  & MHz            & $\mu$Jy/Bm & mJy              & \kmps\     &  mJy    &  \kmps\          & $\times 10^{22}$~\cm &        \\
\hline
J0229+0044 & 02h29m28.9s, & 4.17 &  640.80        & 100	       & $3.43 \pm 0.38$  & 3.8        & 0.87    & $74.2 \pm 7.8$   & $1.35 \pm 0.14$      & 1.2166 \\
       & 00d44$'$29.5$''$ &      &                & 	       &                  &            &         &                  & & \\
J0229+0053 & 02h29m47.2s, & 200  &  656.68        &  20        & $31.08 \pm 0.41$ & 5.6        & 0.41    & $78.41 \pm 0.81$ & $1.429 \pm 0.015$    & 1.1630 \\
       & 00d53$'$08.9$''$ &      &                &            &                  &            &         &                  & & \\
\hline
\end{tabular}
\vskip 0.1in
\end{table*}

Our final \hii\ spectra towards J0229+0044 and J0229+0053 are shown in Figs.~\ref{fig:spectra}[A] and 
\ref{fig:spectra}[B], respectively.  The velocity-integrated \hii\ optical depths are  $(74.2 \pm 7.8)$~\kmps\ (J0229+0044) and $(78.41 \pm 0.81)$~\kmps\ (J0229+0053). Assuming $\ts = 100$~K and a covering factor, $f=1$, these integrated 
\hii\ optical depths imply \hi\ column densities of $\nhi = (1.35 \pm 0.14) 
\times (T_s/100) \times 10^{22}$~cm$^{-2}$ (J0229+0044) and $\nhi = (1.429 \pm 0.015) \times (T_s/100) \times 10^{22}$~cm$^{-2}$ (J0229+0053). Note that these are conservative assumptions: a lower covering factor or a higher spin temperature would imply even higher \hi\ column densities. The observational results are summarized in Table~\ref{table:results}.

{Our correlator setup allowed us to carry out a simultaneous search for the redshifted OH~1665~MHz and 1667~MHz lines towards both sources \citep[e.g.][]{Kanekar02,Kanekar05}. The spectral RMS noise values at the redshifted OH line  frequencies were 0.9~mJy/Bm per 9.7~\kmps\ channel (J0229+0044) and 0.36~mJy/Bm per 9.5~\kmps\ channel (J0229+0053); this yields $3\sigma$ upper limits to the velocity-integrated OH 1667~MHz optical depth of $\approx 9.5$~\kmps\ (J0229+0044) and $\approx 0.25$~\kmps\ (J0229+0053), assuming a Gaussian line profile with a line FWHM of 10~\kmps.}



\section{The AGNs: J0229+0044 and J0229+0053}
\label{sec:results}

\subsection{J0229+0044}
\label{ssec:44}

{The AGN J0229+0044 is spatially coincident with an object identified in the DEEP2 survey (DEEP2~42053345), with a spectroscopic redshift of $z = 1.2161 \pm 0.0045$ from the O{\sc ii}$\lambda 3727$ doublet}\footnote{The object was targeted in the DEEP2 survey at two different epochs and the DEEP2 catalog reports two redshifts, $z=1.21639$ and $z=1.21577$; these measurements are consistent within the $\approx 62$~\kmps\ redshift error for redshift quality Q$=3$ objects \citep[][]{Newman13}. We have used the mean of these two measurements as the AGN redshift, and have assumed the mean estimate to have a $1\sigma$ uncertainty of $\approx 62$~\kmps.}. 
{The quasar has g$=23.43 \pm 0.25$ from the Dark Energy Survey \citep[DES;][]{Abbott18} and K$=19.896 \pm 0.080$ from the ALHAMBRA survey \citep{Moles08}, i.e. g$-$K$=3.53 \pm 0.26$.}  
The object is also detected by GALEX in the near-ultraviolet (NUV) band, with an NUV magnitude of $23.76 \pm 0.22$ (i.e. similar to the g-band magnitude); however, we note that the NUV-band emission may be contaminated by the Lyman-$\alpha$ emission of the AGN. We estimate the rest-frame 1215~\AA\ luminosity\footnote{We assume a flat $\Lambda$-cold dark matter cosmology, with ($\rm H_0$, $\rm \Omega_{m}$, $\rm \Omega_{\Lambda})=(70$~km~s$^{-1}$~Mpc$^{-1}$, $0.3, 0.7)$ to convert flux densities to luminosity densities at the \hii\ absorption redshift.} (L$_{\rm{UV}}$) of J0229+0044 by interpolating between the measured B-band \citep[B$=23.516 \pm 0.058$;][]{Willmer06} 
and NUV magnitudes; this yields a 1215~\AA\ luminosity of L$_{\rm{UV}}=10^{21.68}$~W~Hz$^{-1}$ (this is formally
an upper limit, due to the possibility that the GALEX NUV measurement may be affected by the AGN's
Lyman-$\alpha$ emission).

J0229+0044 has a 641~MHz flux density of  $(3.43\pm 0.38)$~mJy in our GMRT continuum image; this implies 
a rest-frame 1.4~GHz radio luminosity of L$_{\rm 1.4\; GHz}=10^{25.12}$~W~Hz$^{-1}$. Further, the source 
has a flux density of $2.65$~mJy at 1.4~GHz \citep[from the VLA FIRST survey;][]{becker95} and $1.55 \pm 0.07$~mJy 
at 8.5~GHz (from our analysis of an archival VLA X-band data set, project~AC274). Combining the three measurements,
we find that J0229+0044 has a relatively flat radio spectrum, with a spectral index of $\alpha \approx-0.3$ between $641$~MHz and $8.5$~GHz {(where $\alpha$ is defined such that $\textrm{S}_\nu \propto \nu^{\alpha}$)}. 


The \hii\ absorption towards J0229+0044, shown in Figure \ref{fig:spectra}[A], consists of a single component, centered at $z=1.216557 \pm 0.000058$. This is in excellent agreement with the spectroscopic redshift of the DEEP2 object, indicating that the \hii\ absorption arises from gas in the AGN environment. A single-component Gaussian fit to the \hii\ line profile yields a peak optical depth of $\tau_{\rm max} = 0.727 \pm 0.083$ and an FWHM of $56.1 \pm 7.4$~\kmps. 
We emphasize that the optical depth sensitivity of the current GMRT \hii\ spectrum towards J0229+0044 is quite low, due to the low AGN flux density. We hence cannot rule out the presence of the wide associated \hii\ absorption that is typically detected in AGNs with L$_{\rm 1.4\;GHz} > 10^{24}$~W~Hz$^{-1}$ \citep{Maccagni17}.

\subsection{J0229+0053}
\label{ssec:53}

{The \hii\ spectrum towards J0229+0053, shown in Fig.~\ref{fig:spectra}[B], has two distinct features: a 
strong component centred at $z=1.16299$ and with a peak optical depth of $\approx 0.50$, and a wide weak 
wing that extends to $\approx -440$~\kmps\ with respect to the main component. Unfortunately, the AGN does not currently have a spectroscopic redshift; its photometric redshift is $z=0.86 \pm 0.22$ \citep{Hsieh05}.
The redshift of the main \hii\ component is in broad agreement with the AGN's photometric redshift, but the large uncertainty in the photometric redshift means that we cannot formally rule out the possibility that the AGN is a background source \citep[i.e. that the absorption arises from an ``intervening'' galaxy; e.g. ][]{Kanekar14dla}. However, the high velocity-integrated \hii\ optical depth of the absorber suggests that it is likely to be an associated system; such high $\inttau$ values have hitherto only been obtained from \hii\ absorption in AGN environments \citep[e.g.][]{Gereb15,Aditya18cjf,Kanekar14dla}. Further, the wide line profile, with $\Delta \textrm{V}\approx 640$~\kmps\ between the nulls, is significantly larger than those seen in intervening systems, where $\Delta \textrm{V} \lesssim 200$~\kmps\ \citep[][]{Kanekar09}. We hence conclude that the \hii\ absorption in J0229+0053 is likely to arise from gas in the AGN environment, i.e. that the AGN redshift is approximately equal to the \hii\ absorption redshift.}

{J0229+0053 is faint at ultraviolet and optical wavelengths, with V$=24.09 \pm 0.16$, B$=25.23 \pm 0.25$ \citep{Hsieh05}, g=$24.31 \pm 0.29$ \citep[DES;][]{Abbott18}, and a non-detection in the GALEX NUV band. The object is detected in the UKIDSS survey, with K$=20.19 \pm 0.20$ \citep{Lawrence07}.  Interpolating between the B-band magnitude and the upper limit on the NUV magnitude yields the $3\sigma$ upper-limit L$_{\rm{UV}}<10^{21.66}$~W~Hz$^{-1}$ on the 1215~\AA\ luminosity, where we have assumed that the AGN redshift is the same as the \hii\ absorption redshift.}

The measured GMRT 657~MHz flux density of $31.08 \pm 0.16$~mJy implies that J0229+0053 has a 
rest-frame 1.4~GHz luminosity of L$_{\rm 1.4\; GHz}=10^{26.08}$~W~Hz$^{-1}$, {again assuming that the AGN redshift is the same as the \hii\ absorption redshift.} The AGN has a flux density of $\approx 80.4$~mJy at 1.4~GHz \citep[VLA FIRST survey;][]{becker95}, i.e. an inverted spectrum at low frequencies, with spectral index $\alpha\approx 1.3$. The AGN spectral index above 1.4~GHz is uncertain, due to inconsistencies between multiple 4.8~GHz flux-density measurements, $\approx 46$~mJy in the 87GB survey 
\citep{Gregory91} and $\approx 100$~mJy in the MIT Green Bank 5~GHz survey \citep{Bennett86}\footnote{The 
discrepancy in the $4.8$~GHz flux-density measurements could be due to either intrinsic variability or 
uncertainties in the flux density scale of the two surveys.}. These measurements yield spectral indices 
of $\alpha^{1.4 \rm{GHz}}_{4.8 \rm{GHz}} \approx-0.45$ or $=0.18$. Thus, the AGN either is a 
GPS source or has an inverted spectrum at frequencies $\lesssim 5$~GHz.


{While the strength and width of the \hii\ absorption towards J0229+0053 imply that the absorption is likely to arise from gas in the AGN environment, a detailed understanding of the system critically requires a measurement of the AGN redshift. For example, if the AGN redshift is the same as the redshift of the main \hii\ absorption component, the wide and weak \hii\ absorption is likely to trace a jet-driven outflow of cold gas \citep[e.g.][]{Oosterloo00,Morganti13}. Conversely, if the strong \hii\ absorption lies redward of the AGN redshift, the different \hii\ absorption components may arise against distinct sub-arcsecond-scale radio continuum structures \citep[e.g.][]{Peck01,Struve10}.}




\section{Discussion}
\label{sec:discussion}

\begin{figure*}
    \centering
    \includegraphics[width=\linewidth]{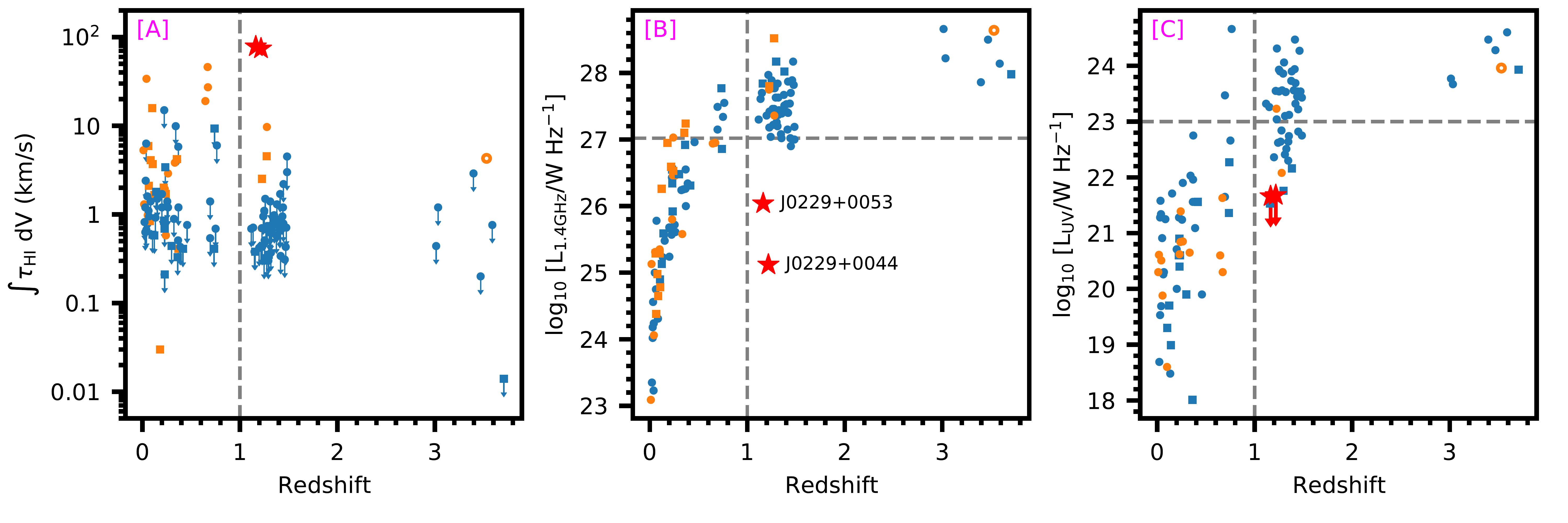}
    \caption{{[A]~The integrated \hii\ optical depth, [B]~the rest-frame 1.4~GHz luminosity (L$_{1.4\rm{GHz}}$), and [C]~the rest-frame 1215~\AA\ UV luminosity (L$_{\rm{UV}}$), plotted against redshift, for flat-spectrum AGNs from the CJF sample \citep[circles;][]{Aditya18cjf}, GPS sources \citep[squares;][]{Aditya18gps}, and the two sources of this {\it Letter} (red stars). Detections of \hii\ absorption in the CJF and GPS samples are shown in orange; for non-detections (shown in blue), the plotted velocity-integrated \hii\ optical depths are $3\sigma$ upper limits, assuming a line FWHM of $100$~\kmps.  The open orange circle is a tentative detection of \hii\ absorption \citep{Aditya16}. The horizontal dashed line in panel~[B] marks the median 
L$_{1.4\rm{GHz}}$ of the CJF and GPS samples, while that in panel~[C] indicates L$_{\rm{UV}}=10^{23}$~W~Hz$^{-1}$.
See main text for discussion.}
    \label{fig:cjf}}
\end{figure*}

{The two new \hii\ absorbers presented here are remarkable due to both the faintness of the background sources, and the high velocity-integrated \hii\ optical depths, larger than that of any known redshifted \hii\ absorber at $z>0.2$ and nearly an order of magnitude higher than those at $z>1$ \citep[e.g.][]{Curran13,Aditya18gps}.} Indeed, this is the first time that \hii\ absorption has ever been detected against a source with an integrated flux density lower than $\approx 25$~mJy at the redshifted \hii\ line frequency at {\it any} redshift \citep[e.g.][]{Gereb15,Maccagni17,Aditya18cjf}; J0229+0044 has a flux density of just $\approx3.4$~mJy! {In passing, we note that VLBI \hii\ spectroscopy of low-$z$ AGNs has yielded detections of \hii\ absorption against resolved source components with individual flux densities $\approx 3-10$~mJy (e.g. \citealp{Jozsa09,Srianand15}), as well as similarly high \hii\ optical depths again against individual source components ($\inttau \approx 50-120$~\kmps; e.g. \citealp{Morganti04,Srianand15}).}


{Both AGNs show evidence for significant reddening, with g$-$K$=3.53$ (J0229+0044) and g$-$K$=4.12$ (J0229+0053). For comparison, only $0.2$\% of the 1,697 optically-selected QSOs at $z = 1.1-1.3$ in the SDSS-UKIDSS matched catalog of \citet{Peth11} have g$-$K$>2$, with none having g$-$K$>2.5$.} The inferred \hi\ column density along the two sightlines is extremely high, $\nhi \approx 1.4 \times (T_s/100) \times 10^{22}$~cm$^{-2}$. This indicates a high dust column along the sightline, which is likely to cause high obscuration, and hence, reddening of the AGN \citep[e.g.][]{Webster95}. Searches for associated \hii\ absorption towards red quasars have shown higher detection rates, consistent with the hypothesis that the reddening is due to dust obscuration in the AGN environment \citep[e.g.][]{Carilli98,Yan16}. {However, it is possible that the apparent reddening might also arise due to the colour of the AGN host galaxy \citep[e.g.][]{Benn98}.} \citet{Aditya18cjf} found no significant evidence of a dependence of the strength of \hii\ absorption on the AGN color, also suggesting that AGN reddening might arise from other causes. 

We emphasize that the absorber towards J0229+0053 would not have been detected if we had limited our search to radio sources with known spectroscopic redshifts. Indeed, the detection of strong \hii\ absorption against J0229+0053, which is highly reddened and hence faint at optical and UV wavelengths, demonstrates the power of such unbiased searches in identifying high \hii\ opacity sightlines that are excellent candidates for molecular absorption.


{\citet{Aditya18gps,Aditya18cjf} used \hii\ absorption surveys of GHz-peaked-spectrum (GPS) AGNs and flat-spectrum AGNs \citep[the latter drawn from the Caltech-Jodrell Flat-spectrum (CJF) sample; e.g. ][]{Taylor96}, respectively, to find that associated \hii\ absorption is significantly weaker in AGNs at both high redshifts, and high rest-frame radio or UV luminosities. These associated \hii\ absorption studies are unique due to the homogeneity of the target AGN samples. Our two AGNs have flat or inverted low-frequency spectra, with radio spectral indices consistent with the spectral-index selection criterion ($\alpha \geq -0.5$) of the CJF sample\footnote{{We note that the spectral index of J0229+0053 is uncertain due to the difference between the two measurements of its 4.8~GHz flux density; however, the two 4.8~GHz measurements yield spectral indices ($\alpha = -0.45, +0.18$) consistent with the CJF selection criterion (see Section \ref{ssec:53}).}}. Fig.~\ref{fig:cjf}[A] shows the velocity-integrated \hii\ optical depths of the 122 \hii\ absorbers of the CJF and GPS samples of \citet{Aditya18gps,Aditya18cjf} plotted as a function of redshift (with detections shown in orange and non-detections in blue). The low \hii\ absorption detection rate and the paucity of high-opacity absorbers at high redshifts are clear from the figure: there are only three confirmed \hii\ absorbers at $z > 1$, all with integrated \hii\ opacities $< 10$~\kmps. Figs.~\ref{fig:cjf}[B] and [C] illustrate the luminosity bias in the samples, with higher-redshift AGNs having higher radio and UV luminosities. The lower strength of associated \hii\ absorption in high-redshift AGNs of the GPS and CJF samples can hence be explained by (1)~redshift evolution of the \hi\ column density or gas spin temperature in AGN environments, (2)~excitation of the upper hyperfine \hi\ level due to the high AGN 1.4~GHz luminosity, resulting in a high spin temperature,  or (3)~ionization of the \hi\ due to the high AGN UV luminosity, resulting in a low \hi\ column density. }

Our two new \hii\ absorbers are shown as red stars in each of the panels of Fig.~\ref{fig:cjf}. It is clear that they have far higher \hii\ optical depths, by nearly an order of magnitude, than any \hii\ absorber from the CJF or GPS samples at $z > 1$. Further, unlike all the CJF and GPS AGNs at $z > 1$, the two AGNs have relatively low luminosities in both the rest-frame 1.4~GHz and the UV wavebands, L$_{\rm 1.4\; GHz} = 10^{25.1} - 10^{26.08}$~W~Hz$^{-1}$ and L$_{\rm{UV}} \leq 10^{21.7}$~W~Hz$^{-1}$. Our discovery of two high-$z$ \hii\ absorbers with high velocity-integrated optical depths, and low rest-frame radio and UV AGN luminosities suggests that ionization and/or excitation effects play an important role in determining the strength of associated \hii\ absorption. It thus appears likely that the current paucity of \hii\ absorbers at $z>1$ may be due to the luminosity bias in AGN samples that have so far been searched for associated \hii\ absorption. 

{Of course, our results do not rule out the possibility of redshift evolution in AGN environments, in either the \hi\ column density or the spin temperature. However, it is clear that tests of putative redshift evolution must be carried out on AGN samples with similar distributions of UV and radio luminosities, across a range of redshifts.}

Finally, we note that the rest-frame 1215~\AA\ luminosities of both J0229+0044 and J0029+0053 are 
significantly lower than the threshold of L$_{\rm{UV}}=10^{23}$~W~Hz$^{-1}$ suggested by \citet{Curran08}, 
above which UV photons from the AGN may ionize most of the \hi. Indeed, all the AGNs that show confirmed 
detections of associated \hii\ absorption at $z > 1$ have low rest-frame UV luminosities, 
L$_{\rm{UV}} \lesssim 2 \times 10^{23}$~W~Hz$^{-1}$ \citep{Curran08,Aditya17,Aditya18gps,Dutta20}, 
but show a wide dispersion in their rest-frame 1.4~GHz 
radio luminosities, L$_{\rm 1.4\; GHz} \approx 10^{25.1} - 10^{28.5}$~W~Hz$^{-1}$. This suggests
that the rarity of associated \hii\ absorbers at $z > 1$ may arise mainly because of a decrease in the \hi\ 
column density in the AGN environment due to ionization of the \hi\ by UV photons, rather than an increase 
in the \hi\ spin temperature due to proximity to the bright radio AGN. 

In summary, we have used an unbiased GMRT \hii\ absorption survey at $z \approx 0.73 - 1.53$ to discover
two new \hii\ absorbers at $z \approx 1.2$, both of which are likely to arise from \hi\ in the AGN 
environment. The two absorbers have the highest integrated \hii\ optical depths of all known redshifted 
\hii\ absorbers, with an inferred \hi\ column density of $\approx 1.4 \times (\ts/100) \times 10^{22}$~\cm. 
Both AGNs are significantly reddened (g$-$K~$=3.53, 4.12$), consistent with dust obscuration at the above high \hi\ column densities. The two AGNs have lower rest-frame 1215~\AA\ UV and 1.4~GHz radio luminosities than most 
high-$z$ AGNs that have hitherto been searched for \hii\ absorption.  Our results suggest that ionization effects 
at high AGN UV luminosities are likely to play an important role in determining the strength of associated \hii\ absorption, and that the current dearth of \hii\ absorbers at $z>1$ may be due to the luminosity bias in current high-$z$ AGN samples with searches for \hii\ absorption. Unbiased \hii\ absorption surveys should allow us to significantly increase the number of \hii\ absorbers at high redshifts in the near future.

\acknowledgments

NK acknowledges support from the Department of Science and Technology via a Swarnajayanti Fellowship (DST/SJF/PSA-01/2012-13). The authors also acknowledge the Department of Atomic Energy for funding support, under project 12-R\&D-TFR-5.02-0700. We thank the staff of the GMRT who have made these observations possible. The GMRT is run by the National Centre for Radio Astrophysics of the Tata Institute of Fundamental Research. It is a great pleasure for A.C. to thank Suma Murthy for numerous insightful discussions that led to the discovery of one of the absorbers reported in this \emph{Letter}. A.C. also thanks Raffaella Morganti for a useful discussion on searching for associated \hii\ absorbers towards sources in the DEEP2 survey. 



\end{document}